\UseRawInputEncoding
\documentclass[aps,
prl,
twocolumn,
longbibliography,
superscriptaddress,
floatfix,
10pt
]{revtex4-2}
\usepackage[utf8]{inputenc}
\usepackage[english]{babel}
\usepackage{amsmath,amssymb,bbm,mathrsfs,bm,braket,color,graphicx,comment,amsfonts,dsfont}
\usepackage[colorlinks,linkcolor=blue,citecolor=blue,urlcolor=blue]{hyperref}
\usepackage[mathscr]{euscript}
\usepackage{physics}
\usepackage{xcolor}
\usepackage[normalem]{ulem}
\usepackage{bm}
\usepackage{orcidlink}
\usepackage{multirow}
\usepackage{microtype}
\usepackage{bbold}
\usepackage{pifont}
\usepackage{slashed}
\usepackage{bibunits}
\usepackage{titlesec}
\usepackage{textcase}
\usepackage{csquotes}
\usepackage{listings}
\definecolor{keyword}{rgb}{0.6,0.2,0.8}  
\definecolor{string}{rgb}{0.2,0.6,0.2}   
\definecolor{comment}{rgb}{0.5,0.5,0.5}  
\definecolor{identifier}{rgb}{0,0,0}     
\definecolor{type}{rgb}{0.2,0.4,0.8}     
\definecolor{operator}{rgb}{0.8,0.1,0.1} 

\lstdefinelanguage{Golang}{
  keywords={break, case, chan, const, continue, default, defer, else, fallthrough, 
            for, func, go, goto, if, import, interface, map, package, range, return, 
            select, struct, switch, type, var},
  keywordstyle=\color{keyword}\bfseries,
  ndkeywords={bool, byte, complex64, complex128, error, float32, float64, int, 
              int8, int16, int32, int64, rune, string, uint, uint8, uint16, 
              uint32, uint64, uintptr, true, false, iota, nil},
  ndkeywordstyle=\color{type}\bfseries,
  identifierstyle=\color{identifier},
  sensitive=true,
  comment=[l]{//},
  morecomment=[s]{/*}{*/},
  commentstyle=\color{comment}\itshape,
  stringstyle=\color{string}\ttfamily,
  morestring=[b]',
  morestring=[b]"
}

\lstdefinestyle{customc}{
  belowcaptionskip=1\baselineskip,
  breaklines=false,
  frame=L,
  xleftmargin=\parindent,
  language=C,
  showstringspaces=false,
  basicstyle=\footnotesize\ttfamily,
  keywordstyle=\bfseries\color{green!40!black},
  commentstyle=\itshape\color{purple!40!black},
  identifierstyle=\color{blue},
  stringstyle=\color{red},
}

\begin{document}
\title{Three-Dimensional Shankar Skyrmions in Frustrated Antiferromagnets}
\author{Vladyslav M. Kuchkin}
\email{vladyslav.kuchkin@uni.lu}
\affiliation{Department of Physics and Materials Science, University of Luxembourg, L-1511 Luxembourg, Luxembourg}
\author{Ricardo Rama-Eiroa}
\email{ricardo.rama-eiroa@ed.ac.uk}
\affiliation{Institute for Condensed Matter and Complex Systems, School of Physics and Astronomy, University of Edinburgh, Edinburgh, United Kingdom}
\affiliation{Higgs Centre for Theoretical Physics, The University of Edinburgh, Edinburgh, United Kingdom}
\author{Carlos Saji}
\affiliation{Departamento de F\'isica, CEDENNA, FCFM, Universidad de Chile, Santiago, Chile}
\author{Alvaro S. Nunez}
\affiliation{Departamento de F\'isica, CEDENNA, FCFM, Universidad de Chile, Santiago, Chile}
\author{Roberto E. Troncoso}
\email{r.troncoso.c@gmail.com}
\affiliation{Instituto de Alta Investigaci\'on, Universidad de Tarapac\'a, Casilla 7D, Arica, Chile}

\begin{abstract}
We formulate a continuum theory for three-dimensional (3D) Shankar skyrmions in frustrated chiral antiferromagnets (AFs) and derive the conditions for metastable finite-size $\pi_3(SO(3))$ solitons. A Derrick--Hobart scaling analysis shows that exchange, Dzyaloshinskii--Moriya interaction (DMI), anisotropy, and frustration can balance to set a finite equilibrium size. We identify two microscopic routes to this common topological and stabilization framework: intrinsic rotation-frame order in noncollinear AFs and an $\mathbb{S}^3$ extension of an amplitude-softened Néel field. In the latter, the smooth four-component texture projects onto a Néel field containing a spatially separated pair of oppositely charged Bloch points. Numerical minimization yields metastable monopole textures, while the effective dynamics identifies coherent breathing oscillations as their characteristic finite-frequency collective mode. Our results provide a microscopic framework for the statics and dynamics of Shankar skyrmions and identify frustrated chiral AFs as promising hosts of 3D non-Abelian topological textures.
\end{abstract}

\maketitle
Topological magnetic solitons are spatially localized magnetization textures that exhibit rich topological phases and unconventional dynamics \cite{NagaosaTokura,Gbel2021,Kuchkin2025_v3}. Their stability and mobility make them promising building blocks for high-density information storage and processing. Chiral domain walls \cite{Thiaville2012} and skyrmions \cite{BogdanovYablonskii,Muhlbauer2009,Yu2010,Brearton2022} are prominent examples of low-dimensional textures and model systems for realization of emergent electrodynamics and protected topological dynamics \cite{NagaosaTokura}.

3D topological textures emerge from richer order-parameter manifolds {in various physical systems~\cite{Volovik2020, Chen2013, Tai2019, Tai2022, Hall2025, Lukyanchuk2020, GmezOrtiz2024}} and support excitations with no direct lower-dimensional analogue. 
{In magnetism,} representative examples include Bloch points (BPs), which are singular point defects~\cite{Tapia2024, CarvalhoSantos2015, Elias2014, Kuchkin2025_v2}, and hopfions, smooth textures associated with maps $\mathbb{S}^3\!\to\!\mathbb{S}^2$ and classified by the Hopf invariant~\cite{FaddeevNiemi,Rybakov2022,Saji2023-prl}. A distinct class of nonsingular 3D solitons arises when the order parameter is a rotation field, with topology classified by $\pi_3(SO(3))=\mathbb Z$. Its paradigmatic representative is the Shankar skyrmion, or Shankar monopole, originally introduced as a topological excitation in ordered media~\cite{ShankarPRD,Shankar1977}. The texture is described by a rotation field ${\cal R}(\mathbf r)\in SO(3)$, whose integer winding is quantified by~\cite{Volovik1977ParticleLike, Polyakov1987}
\begin{align}\label{eq: topological-charge}
Q=\frac{1}{96\pi^2}\int d^3x\,\epsilon^{ijk}\,
\operatorname{Tr}\!\left[{\cal L}_i{\cal L}_j{\cal L}_k\right],
\end{align}
where ${\cal L}_i={\cal R}^{-1}\partial_i{\cal R}$ is the $\mathfrak{so}(3)$-valued Maurer--Cartan field \footnote{Eq. \eqref{eq: topological-charge} is the $SO(3)$ counterpart of Polyakov's
$SU(2)$ winding formula~\cite{Polyakov1987}. Let $g\in SU(2)$ be a
lift of $R\in SO(3)$ under the double covering
$SU(2)\to SO(3)$, and define $\ell_i=g^{-1}\partial_i g$. The traces
in the adjoint and fundamental representations satisfy
$\operatorname{Tr}_3(\mathcal L_i\mathcal L_j\mathcal L_k)
=4\operatorname{tr}_2(\ell_i\ell_j\ell_k)$. Consequently,
Eq. \eqref{eq: topological-charge} is equivalent to
$Q=(24\pi^2)^{-1}\int d^3x\,\epsilon^{ijk}
\operatorname{tr}_2(\ell_i\ell_j\ell_k)$.
Here $\operatorname{Tr}$ in Eq. \eqref{eq: topological-charge} denotes the ordinary $3\times3$ matrix trace, and $\epsilon^{123}=+1$.}. A representative of the unit-winding sector is the Shankar ansatz~\cite{Shankar1977}, ${\cal R}_{S}(\mathbf r)=\exp\,[ f(r)\, \hat{\mathbf r}\!\cdot\!\hat{\mathbf L}]$, where $\hat{\mathbf r}=\mathbf r/r$ and $\hat{\mathbf L}=(\hat L_x,\hat L_y,\hat L_z)$ are the generators of $\mathfrak{so}(3)$. The boundary conditions $f(0)=2\pi$ and $f(r\!\to\!\infty)=0$ imply ${\cal R}_{S}(0)={\cal R}_{S}(\infty)=\mathbbm{1}$. The fixed asymptotic value compactifies physical space as ${\mathbb S}^3_{\rm space}$, so that ${\cal R}_{S}$ defines a smooth map $\mathbb{S}^3_{\rm space}\rightarrow SO(3)$ with $Q=1$. This unit-winding construction provides the common topological reference for the microscopic realizations of Shankar skyrmions. Despite its long-standing theoretical foundation, their realization in magnetic materials has remained elusive, as finite-size stabilization requires competing interactions beyond the scale-invariant nonlinear sigma model~\cite{Shankar1977}. Related work has identified 3D topological textures~\cite{barts2021magnetic} and non-Abelian spin dynamics~\cite{zarzuela2019hydrodynamics,Zarzuela2025}, but a microscopic understanding of the existence, stabilization, and dynamics of Shankar skyrmions remains lacking.

In this Letter, we establish a framework to realize 3D Shankar skyrmions in frustrated AFs. Frustrated AFs provide two natural microscopic routes: in bipartite systems the order parameter can acquire an amplitude-softened $\mathbb{S}^3$ structure, whereas in noncollinear systems the low-energy manifold is described directly by a local $SO(3)$ rotation field. We show that exchange, DMI, anisotropy, and frustration provide the competing energy scales required to stabilize finite-size textures with integer Shankar winding. Numerical minimization yields metastable monopole textures, while their low-energy dynamics exhibits coherent breathing oscillations as a characteristic finite-frequency collective excitation. Together, these results unify stabilization and microscopic realization of Shankar skyrmions, establishing frustrated AFs as a platform for 3D non-Abelian magnetism.
\begin{figure*}[!htb]
\centering
\includegraphics[width=18cm]{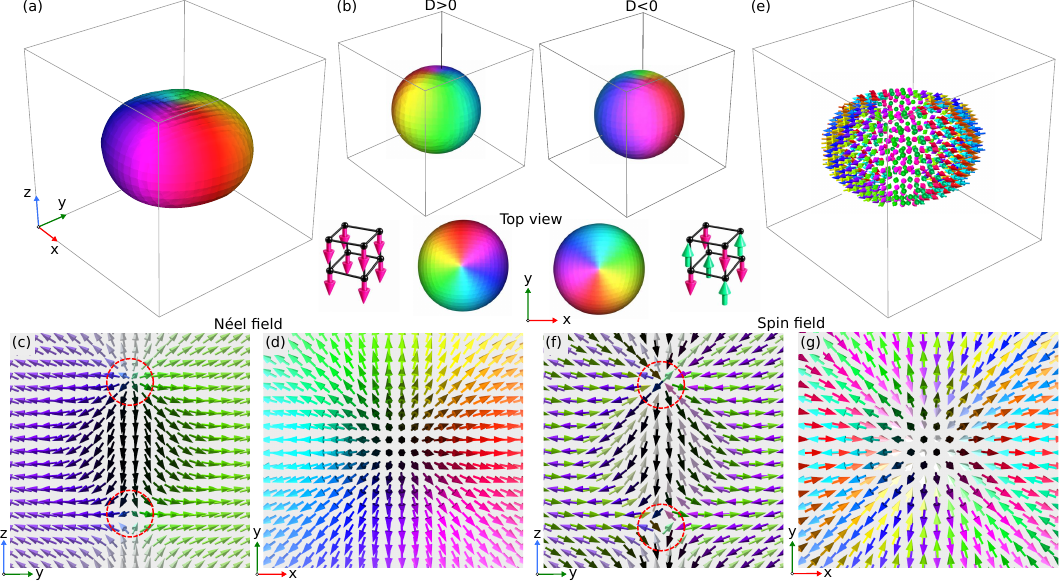}
\caption{Numerically relaxed 3D Shankar texture in a frustrated bipartite AF. (a) Staggered Néel field $\mathbf{n}$ in the full simulation cell. (b) Néel-field configurations stabilized for $D>0$ and $D<0$; the lower images provide top views and highlight the change in color winding upon reversing the sign of $D$. Panels (c) and (d) show the central longitudinal $yz$ ($x=0$) and transverse $xy$ ($z=0$) cross sections of the Néel texture. (e) Corresponding microscopic spin field $\mathbf{s}$, with the analogous cross sections shown in (f) and (g). Arrows and colors encode the local orientation of the respective vector fields. The lattice schematics illustrate the staggered transformation on the two AF sublattices. The relaxation was performed on a $64^{3}$ cubic lattice with periodic boundary conditions, using $J_{4}=0.05$ and $J_{1}=0.2$.
The longitudinal sections in (c) and (f) reveal a finite skyrmion-tube segment terminated by two BPs, whose positions are marked by dashed circles, whereas the transverse sections in (d) and (g) display the associated in-plane winding. In the Néel description, these endpoints coincide with zeros of the non-normalized staggered field, at which the normalized field $\hat{\mathbf{n}}$ becomes ill-defined.}
\label{Figure1:Shankar-solutions}
\end{figure*}

As a first microscopic realization, we consider a frustrated Heisenberg AF, $H_1=\sum_{ij}J_{ij}\mathbf S_i\cdot\mathbf S_j$, with $J_{ij}$ the exchange interactions between localized spins at $i$ and $j$ positions. For a bipartite AF, the low-energy order parameter is the Néel field $\mathbf n$, with $|\mathbf n|=1$ in the ordered phase, and the lattice Hamiltonian becomes $H_1=\sum_{ij}J_{ij}^{\prime}\mathbf n_i\cdot\mathbf n_j+{\cal O}[\mathbf m]$. For the first four exchange shells, $r_1=a$, $r_2=a\sqrt{2}$, $r_3=a\sqrt{3}$, and $r_4=2a$, the staggered transformation gives $J_1^{\prime}=-J_1$, $J_2^{\prime}=J_2$, $J_3^{\prime}=-J_3$, and $J_4^{\prime}=J_4$. For slowly varying textures, a gradient expansion yields a frustrated continuum functional $\mathcal H[\mathbf n]$, whose higher-order gradient terms originate from competing exchange interactions~\cite{Rybakov2022}:
{
\begin{equation}
        \mathcal H[\mathbf n]\!\!=\!\!\int\!\!\mathrm{d}^{3}x\!\left[\mathcal{A}\left({\partial_{\alpha}\mathbf{n}}\right)^{2}\!+\!\mathcal{B}\left({\partial^{2}_{\alpha}\mathbf{n}}\!-\!{\partial^{2}_{\beta}\mathbf{n}}\right)^{2} \!+\! \mathcal{C}\!\left({\partial^{2}_{\alpha\beta}\mathbf{n}}\right)^{2}\right]\!\!,\!\!\label{Micro_S2}
\end{equation}
where summation over non-repeating indices $\alpha, \beta \in \left\{x,y,z\right\}$ is assumed; $\mathcal{A}$ is Heisenberg exchange and $\mathcal{B}$, $\mathcal{C}$ are high-order exchange constants. To be consistent with the above atomistic model, we fix $\mathcal{C}=2\mathcal{B}$ below.
}
This fixed-length theory is formally equivalent to the corresponding frustrated-ferromagnet continuum model and
therefore supports the same family of smooth frustration-stabilized textures, including hopfions~\cite{Saji2023-prl}, skyrmion tubes~\cite{Wolf2021}, dipole strings \cite{Kuchkin2025_v1}, and twisted skyrmion tubes~\cite{Saji2025-twisted}.

At a BP, however, a fixed-length N\'eel field cannot remain smooth: its orientation becomes ill defined and the higher-order gradient terms diverge~\cite{Kuchkin2026}. We therefore relax the constraint $|\mathbf n|=1$ and allow the N\'eel amplitude to soften, $|\mathbf n|\leq1$. The field is then embedded into a unit four-component vector $\boldsymbol{\nu}=(\nu_1,\nu_2,\nu_3,\nu_4)\in \mathbb{S}^3$, defined by $\nu_\alpha=n_\alpha$ and $\nu_4^2=1-|\mathbf n|^2$~\cite{Kuchkin2026}.
The additional component allows the local AF order to soften continuously to $|\mathbf n|=0$.
The continuum functional ~\eqref{Micro_S2} generalizes to
\begin{align}
\mathcal{H}^{*}&[\boldsymbol{\nu}] = \mathcal H[\bm{\nu}]+ \kappa\int\!\!\mathrm{d}^{3}x\nu_{4}^{2},
\label{Micro_S3}
\end{align}
where $\kappa>0$ selects the AF vacuum $\nu_4=0$ and $|\mathbf n|=1$. The component $\nu_4$ is allowed to take both signs, so that $\boldsymbol{\nu}$ explores the full space $\mathbb{S}^3$.
{Note that all spatial derivatives in \eqref{Micro_S2} act on vector $\bm{\nu}$ in $\mathcal{H}^{*}[\boldsymbol{\nu}]$ case.}
For a localized texture we impose the fixed asymptotic condition $\boldsymbol{\nu}(\mathbf r)\xrightarrow[]{r\to\infty} \boldsymbol{\nu}_{\infty}=(\mathbf n_{\infty},0),$ and $|\mathbf n_{\infty}|=1$, so that spatial infinity is mapped to a single point of the vacuum manifold. Physical space is then compactified to ${\mathbb{S}}^3_{\rm space}$, and $\boldsymbol{\nu}:{\mathbb{S}}^3_{\rm space}\rightarrow {\mathbb{S}}^3$ is classified by an integer winding number $Q_\nu\in\pi_3(\mathbb{S}^3)\simeq\mathbb Z$, therefore this winding belongs to the same homotopy class as the Shankar charge defined in Eq.~(\ref{eq: topological-charge}). 

\begin{figure}
\centering
\includegraphics[width=8cm]{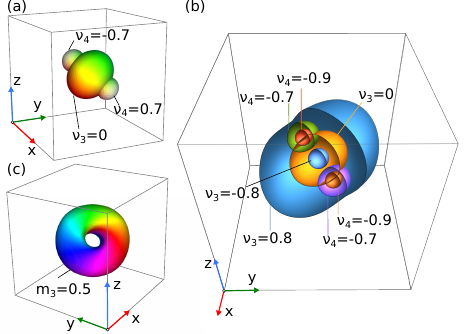}
\caption{Numerically stabilized Shankar monopole in the model of Eq.~\eqref{Micro_S3}. (a) The four-component field $\boldsymbol{\nu}$ is represented by the isosurfaces $\nu_{3}=0$ and $\nu_{4}=\pm0.7$. (b) Additional level sets, $\nu_{3}=0,\pm0.8$ and $\nu_{4}=\pm0.7, \pm0.9$, reveal the nested and mutually linked structure associated with the nontrivial three-dimensional winding. (c) Under the Hopf map,
$m_{1}=2(\nu_{1}\nu_{3}+\nu_{2}\nu_{4})$,
$m_{2}=2(\nu_{2}\nu_{3}-\nu_{1}\nu_{4})$, and
$m_{3}=\nu_{1}^{2}+\nu_{2}^{2}-\nu_{3}^{2}-\nu_{4}^{2}$,
the Shankar texture is projected onto a smooth hopfion, shown through the isosurface $m_{3}=0$, with color encoding the in-plane components $(m_{1},m_{2})$.}
\label{Fig2}
\end{figure}

Fig.~\ref{Figure1:Shankar-solutions} shows the relaxed Shankar texture in both the staggered N\'eel (panel (a)) and microscopic spin (panel (e)) representations. 
The texture contains two antiferromagnetic BPs ~\cite{Kuchkin2025_v1,Kuchkin2025_v2}, identified in the central cross sections ($x=0$) of the N\'eel and spin field, panels (c) and (f) respectively. These singular cores are regularized by the smooth four-component field $\boldsymbol{\nu}$, which carries the nonzero Shankar winding. Fig.~\ref{Figure1:Shankar-solutions}(b) and the corresponding $n_{z}=0$ isosurfaces, further show that reversing the sign of the DMI reverses the handedness of the texture \footnote{A conventional Dzyaloshinskii--Moriya term,
$H_D=\sum_{ij}\mathbf D_{ij}\cdot(\mathbf S_i\times\mathbf S_j)$,
is added to the spin Hamiltonian $H_1$.}. Fig.~\ref{Figure1:Shankar-solutions}(e) shows discrete-lattice Shankar textures, and cross sections $yz\,(x=0)$ and $xy\,(z=0)$ at panels (f) and (g), obtained without DMI. Thus, frustration provides the finite-size stabilization, whereas the DMI selects its chirality of the resulting Shankar texture. The topology of the underlying $\mathbb{S}^{3}$ field is made explicit in Fig.~\ref{Fig2}. Fig.~\ref{Fig2}(a,b) show the unit-winding solution of Eq.~\eqref{Micro_S3}, visualized
through various isosurfaces given by fixed values $\nu_3$ and $\nu_4$.
Applying the Hopf map $\boldsymbol{\nu}\mapsto\mathbf m\in \mathbb{S}^2$ yields Fig.~\ref{Fig2}(c), where the same configuration with $Q_\nu=1$ is represented as a Hopf texture with $H=1$ (the Hopf index ~\cite{Whitehead1947}) \footnote{The Hopf index is $H=-\int_V \mathrm{d}^3 r\, \mathbf{F}\cdot\mathbf{A},$ where the components of the emergent field $\mathbf{F}$ are  $F_i=\epsilon_{ijk}\,\mathbf{m}\cdot\left(\partial_j\mathbf{m}\times\partial_k\mathbf{m}\right)/8\pi,$ with $i,j,k\in\{x,y,z\}$, and $\mathbf{A}$ is the corresponding gauge potential, defined through $\mathbf{F}=\boldsymbol{\nabla}\times\mathbf{A}$}. This makes explicit that the BP structure belongs to the N\'eel projection, whereas the full four-component field remains smooth and topologically nontrivial. For comparison, the fixed-length N\'eel theory also supports smooth frustration-stabilized textures without BP cores, such as the hopfion shown in Fig.~\ref{Figure4: hopfion-solution} at Supplemental Material (SM).

A second route to stabilize Shankar skyrmions is provided by noncollinear AFs with tetrahedral reference order, relevant to pyrochlore magnets~\cite{Gardner2010}. We consider the spin Hamiltonian $H_2=\sum_{\langle i,j\rangle} [J\,\mathbf S_i\cdot\mathbf S_j +\mathbf D_{ij}\cdot(\mathbf S_i\times\mathbf S_j)] -K\sum_i(\mathbf S_i\cdot\mathbf e_i)^2$, where $J>0$ is the antiferromagnetic exchange, $\mathbf D_{ij}$ is the DMI vector, and $\mathbf e_i$ denotes the local easy axis. Within the rigid-frame approximation, smooth rotations of the reference state
are parametrized as $\mathbf S(\mathbf R_{i,\mu}) =\mathcal R(\mathbf R_i)\mathbf e_\mu$, with $\mu=0,1,2,3$. This restriction retains the orientation of the noncollinear frame while keeping the relative sublattice angles fixed~\cite{zarzuela2019hydrodynamics}. For spin variations on length scales much larger than the lattice spacing, a gradient expansion gives the continuum energy $E[\mathcal R]=\int d^3r\,\mathcal E_{\rm tot}$. Retaining even exchange gradients through fourth order and DMI through first order yields
\begin{align}
\mathcal E_{\rm tot}=&\operatorname{Tr}\!\left[G^{ij}\partial_i\mathcal R^T\partial_j\mathcal R+H^i(\mathcal R)\mathcal L_i \right]+K U(\mathcal R) 
\nonumber\\
&\qquad\qquad+\operatorname{Tr}\!\left[\mathsf W^{ijkl}(\partial_i\partial_j\mathcal R)^T \partial_k\partial_l\mathcal R \right],
\label{eq:continuum-energy}
\end{align}
where the constant exchange energy is omitted. The tensors $G^{ij}$ and $\mathsf W^{ijkl}$ are fixed by the second and fourth spatial moments of the exchange, while $H^i(\mathcal R)$ follows from the first moments of the microscopic DMI vectors. This continuum truncation assumes vanishing odd exchange moments and neglects higher-gradient DMI contributions. The gradient-linear term $\operatorname{Tr}[H^i(\mathcal R)\mathcal L_i]$ is a Lifshitz contribution when allowed by the microscopic symmetries~\cite{Elhajal2005}. The single-ion potential is $U(\mathcal R)=-v_c^{-1}\sum_\mu [\operatorname{Tr}(\Pi_\mu\mathcal R)]^2$, where $\Pi_\mu=\mathbf e_\mu\mathbf e_\mu^T$ and $v_c$ is the cell volume.

The finite-size stability criterion follows directly from Eq.~\eqref{eq:continuum-energy} by Derrick--Hobart scaling~\cite{Derrick1964,Hobart1963}. For homogeneous coefficients and a localized reference texture $\mathcal R_0$ approaching a fixed $\mathcal R_\infty$, the dilation $\mathcal R_\lambda(\mathbf r) =\mathcal R_0(\mathbf r/\lambda)$ gives the excess energy $E_{\rm exc}(\lambda)=C_J\lambda+C_D\lambda^2+C_K\lambda^3+{C_W}/{\lambda}.$ Here $C_J$, $C_D$, $C_K$, and $C_W$ are the integrated two-gradient exchange, DMI, anisotropy, and four-gradient contributions evaluated on $\mathcal R_0$ in Eq.~\eqref{eq:continuum-energy} and given at SM. The scaling exponents depend only on derivative order and remain unchanged by tensor anisotropy. Choosing the stationary scale as $\lambda=1$ yields $C_J+2C_D+3C_K-C_W=0$, while stability against uniform dilations requires $E_{\rm exc}''(1)=2(C_D+3C_K+C_W)>0$. For $C_W>0$, the four-gradient energy opposes uniform contraction. In the exchange--frustration limit $C_D=C_K=0$, positive $C_J$ and $C_W$ give $\lambda_\star=\sqrt{C_W/C_J}$, demonstrating finite-size stabilization within this scaling family without DMI~\cite{Skyrme1961,Rybakov2022}, where the required signs follow from the microscopic exchange moments.

The two microscopic realizations carry the same global Shankar topology while resolving the soliton core through different local fields, similar to $SU(2)$ skyrmions, coreless Mermin--Ho and Anderson--Toulouse vortex textures~\cite{Skyrme1961,Mermin1976,Anderson1977}. Firstly, in the amplitude-softened realization, $\boldsymbol{\nu}$ is a smooth four-component field, and for a nonzero winding $Q_\nu$, the field covers the full manifold $\mathbb{S}^3$, including the two poles $N_{\pm}=(0,0,0,\pm 1)$. At the spatial points mapped to these poles, $\mathbf n=0$, so that the normalized N\'eel field $\hat{\mathbf n}=\mathbf n/|\mathbf n|$ becomes undefined, which correspond to the BP cores. For the configuration with $Q_\nu=1$, each pole has a single preimage, giving the two BPs observed in Fig.~\ref{Figure1:Shankar-solutions}. Their local charges~\cite{Malozemoff_79},
\begin{equation}
q_a = \frac{1}{4\pi}\oint_{S_a^{2}}
d\theta\,d\phi\,\hat{\mathbf n}\cdot
\left(\partial_{\theta}\hat{\mathbf n}\times
\partial_{\phi}\hat{\mathbf n}\right),
\end{equation}
satisfy $q_1=-q_2=1$ and characterize the local singularities. The resulting hierarchy therefore separates the global Shankar topology from its morphology of BPs, where the two cores are paired and cannot be removed individually without changing $Q_{\nu}$. Secondly, the rotation-field realization is nonsingular at the level of the full order parameter, $\mathcal R(\mathbf r)$, since it preserves the norm of any nonzero vector. Consequently, every axis of the local spin frame, $\mathbf e_\mu(\mathbf r)=\mathcal R(\mathbf r) \mathbf e_\mu^{(0)}$, remains well defined and non-vanishing throughout space. In particular, the unit-vector projection $\mathbf E_3(\mathbf r)=\mathcal R(\mathbf r)\hat{\mathbf z}$ has $|\mathbf E_3|=1$ identically, so the Shankar winding is revealed through the linking of its regular preimages, without developing BPs.

The dynamics of the Shankar monopole is governed by fluctuations about its equilibrium texture, ${\cal R}_{S}(\mathbf r)$. It comprises core-localized magnons, extended spin waves, and collective translational, rotational, and breathing modes. We probe the dynamics through atomistic spin-dynamics simulations of the frustrated Heisenberg AF $H_1$ using Magnoom~\cite{savchenko2022chiral}. Starting from the relaxed texture with Gilbert damping $\alpha=10^{-4}$, we apply a weak rectangular easy-axis anisotropy pulse of amplitude $K_z=10^{-5}$ and duration $t_{\mathrm p}=1$. This broadband perturbation excites the linear-response spectrum without selecting a frequency. After the pulse, we monitor the free evolution and compare it with an
otherwise identical undriven trajectory to distinguish the induced response from residual relaxation.

\begin{figure}
\centering
\includegraphics[width=8cm]{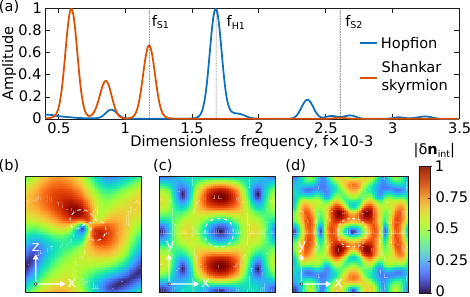}
\caption{Low-amplitude dynamics of the hopfion and Shankar skyrmion. (a) Normalized spectral response following the weak easy-axis anisotropy pulse. The labeled resonances $f_{\mathrm{H1}}$, $f_{\mathrm{S1}}$, and $f_{\mathrm{S2}}$ denote the hopfion internal resonance and the two texture-associated Shankar resonances, respectively. (b--d) Spatial profiles of the normalized internal mode amplitude, $\left| \delta \mathbf{n}_{\rm{int}} \left( \mathbf{r}; f \right) \right|$, where $\delta \mathbf{n}_{\rm{int}} \left( \mathbf{r}; f \right)$ is the local harmonic
response at frequency $f$ after subtraction of the best-fit global rotation. (b) Hopfion at $f_{\mathrm{H1}} \simeq 1.68 \times 10^{-3}$, shown in the central
$xz$ cross-section ($y=33$). (c,d) Shankar skyrmion at $f_{\mathrm{S1}} \simeq 1.18 \times 10^{-3}$ and
$f_{\mathrm{S2}} \simeq 2.61 \times 10^{-3}$, respectively, shown in the central $xy$ cross-section ($z=33$). Here, $\mathbf{n}$ is the N\'eel vector and $\mathbf{r}$ the dimensionless spatial coordinate. The white dashed contours indicate $n^z_0=0$ for the corresponding reference textures and are shown as spatial guides. Each spatial map is normalized independently to its maximum.}
\label{FFT}
\end{figure}

Fig.~\ref{FFT} compares the low-amplitude dynamical response of the Shankar skyrmion with that of a smooth AF hopfion under the same weak broadband driving. Their spectra, shown in Fig.~\ref{FFT}(a), reveal a hopfion resonance at $f_{\mathrm{H}1} \simeq 1.68 \times 10^{-3}$ and a multimodal Shankar response, with a dominant texture-associated resonance at $f_{\mathrm{S}1} \simeq 1.18 \times 10^{-3}$ and a weaker one at $f_{\mathrm{S}2} \simeq 2.61 \times 10^{-3}$, in addition to low-frequency extended excitations. Not every spectral peak can be identified with a texture-associated internal mode; we therefore focus on the resonances whose spatial response is associated with the textures, as discussed in SM Sec.~\ref{sec:sm-mode-identification}. The hopfion remains approximately centered and displays a smooth internal deformation without singular cores at $f_{\mathrm{H}1}$ (Fig.~\ref{FFT}(b)). By contrast, the Shankar texture remains localized, with the finite skyrmion-tube segment and its terminating BPs effectively pinned to the lattice, so that the pulse excites predominantly non-rigid internal deformations rather than translation or global rotation. Its dominant resonance exhibits a spatial structure centered on the Shankar texture (Fig.~\ref{FFT}(c)) and coherent breathing-like dynamics, consistent with excitation of the size coordinate $\lambda(t)$, whereas $f_{\mathrm{S2}}$ represents a distinct internal mode (Fig.~\ref{FFT}(d)). Such multimode behavior is consistent with previous hopfion spectroscopy, where breathing and rotational modes can hybridize~\cite{bo2021spin} and spatially localized breathing excitations have been reported~\cite{tejo2026sub}. Thus, weak broadband driving reveals predominantly internal dynamics in both textures, distinguished by a smooth hopfion deformation and a pinned, breathing-like Shankar response.

We have established a unified continuum framework for finite-size Shankar skyrmions in frustrated chiral AFs. The $SO(3)$ rotation-field theory fixes their $\pi_3(SO(3))$ topology and Derrick--Hobart stability criterion, while two complementary microscopic routes realize this structure: an amplitude-softened $S^3$ embedding of collinear N\'eel order and intrinsic rotation-frame order in noncollinear AFs. Both constructions support an integer-valued 3D winding and share the same Derrick-scaling sectors. The two-gradient stiffness, local-potential, and frustration-induced four-gradient terms scale as $\lambda$, $\lambda^3$, and $\lambda^{-1}$, respectively, and their competition sets a finite equilibrium size; when present, the DMI selects the texture chirality. Numerical minimization confirms metastable monopole solutions, while the effective dynamics separates collective, core-localized-magnon, and continuum excitations and identifies coherent breathing oscillations as a characteristic finite-frequency collective mode. This unified hierarchy connects the microscopic origin, stabilization, and dynamics of Shankar textures, establishing frustrated AFs as a promising platform for 3D non-Abelian topological magnetism.

{\it Acknowledgements} Funding is acknowledged from Fondecyt Regular 1230747 and 1230515, and ANID CEDENNA CIA 250002. R.R.-E. acknowledges financial support from the Royal Society through the Newton International Fellowship NIF\textbackslash R1\textbackslash 241532, "Antiferromagnetic all-spintronics computing based on emergent electrodynamics", and the computational resources provided by the Edinburgh Compute and Data Facility (ECDF) (\href{http://www.ecdf.ed.ac.uk/}{http://www.ecdf.ed.ac.uk}). V.M.K. acknowledges the financial support from the European Union’s Horizon Europe research and innovation programme under the Marie Sk{\l}odowska-Curie grant agreement No.~101203692 (QUANTHOPF) and from the Luxembourg National Research Fund under Grants C22/MS/17415246/DeQuSky and AFR/23/17951349.

{\it Author contributions}
C.S., R.E.T. and A.S.N., conceived the project and carried out the initial analysis.
V.M.K. performed the micromagnetic simulations for stability estimation. R.R.E. performed atomistic spin dynamics and calculated the corresponding spectra.
R.E.T. wrote the first draft. R.E.T. and, together with A.S.N., supervised the project.
All authors interpreted the results and contributed to writing and revising the manuscript.

\bibliographystyle{apsrev4-2}
\bibliography{Bib-3D-Skyrmions}

\newpage
\clearpage
\onecolumngrid

\setcounter{section}{0}
\setcounter{subsection}{0}
\setcounter{secnumdepth}{2}

\renewcommand{\thesection}{S\arabic{section}}
\renewcommand{\thesubsection}{\thesection.\arabic{subsection}}

\makeatletter
\renewcommand{\p@subsection}{}
\makeatother

\makeatletter

\renewcommand{\section}{%
  \@startsection{section}{1}{0pt}%
  {2.0ex plus 0.5ex minus 0.2ex}%
  {0.8ex plus 0.2ex}%
  {\normalfont\normalsize\bfseries\raggedright}%
}

\renewcommand{\subsection}{%
  \@startsection{subsection}{2}{0pt}%
  {1.5ex plus 0.4ex minus 0.2ex}%
  {0.6ex plus 0.2ex}%
  {\normalfont\normalsize\bfseries\raggedright}%
}

\makeatother

\begin{bibunit}[apsrev4-2]

\makeatletter
\renewcommand{\@biblabel}[1]{[S#1]}
\makeatother

\renewcommand{\citenumfont}[1]{S#1}

\setcounter{page}{1}

\setcounter{figure}{0}
\renewcommand{\thefigure}{S\arabic{figure}}

\setcounter{equation}{0}
\renewcommand{\theequation}{S\arabic{equation}}

\setcounter{table}{0}
\renewcommand{\thetable}{S\arabic{table}}

\begin{center}
{\large\bfseries
Supplementary Material for ``Three-Dimensional Shankar Skyrmions in
Frustrated Antiferromagnets''}

\vspace{0.5em}

Vladyslav M. Kuchkin$^{1,*}$,
Ricardo Rama-Eiroa$^{2,3,\dagger}$,
Carlos Saji$^{4}$,
Alvaro S. Nunez$^{4}$,
and Roberto E. Troncoso$^{5,\ddagger}$

\vspace{0.3em}

{\small
$^1$Department of Physics and Materials Science, University of Luxembourg,
L-1511 Luxembourg, Luxembourg\\
$^2$Institute for Condensed Matter and Complex Systems,
School of Physics and Astronomy, University of Edinburgh, \\
Edinburgh, United Kingdom\\
$^3$Higgs Centre for Theoretical Physics, The University of Edinburgh, Edinburgh, United Kingdom\\
$^4$Departamento de Física, CEDENNA, FCFM, Universidad de Chile,
Santiago, Chile\\
$^5$Instituto de Alta Investigación, Universidad de Tarapacá,
Casilla 7D, Arica, Chile
}
\end{center}

\section{\NoCaseChange{Hopfion texture in the frustrated AF}}\label{sec:sm-sm-hopfion}

\begin{figure}[htb]
\centering
\includegraphics[width=8cm]{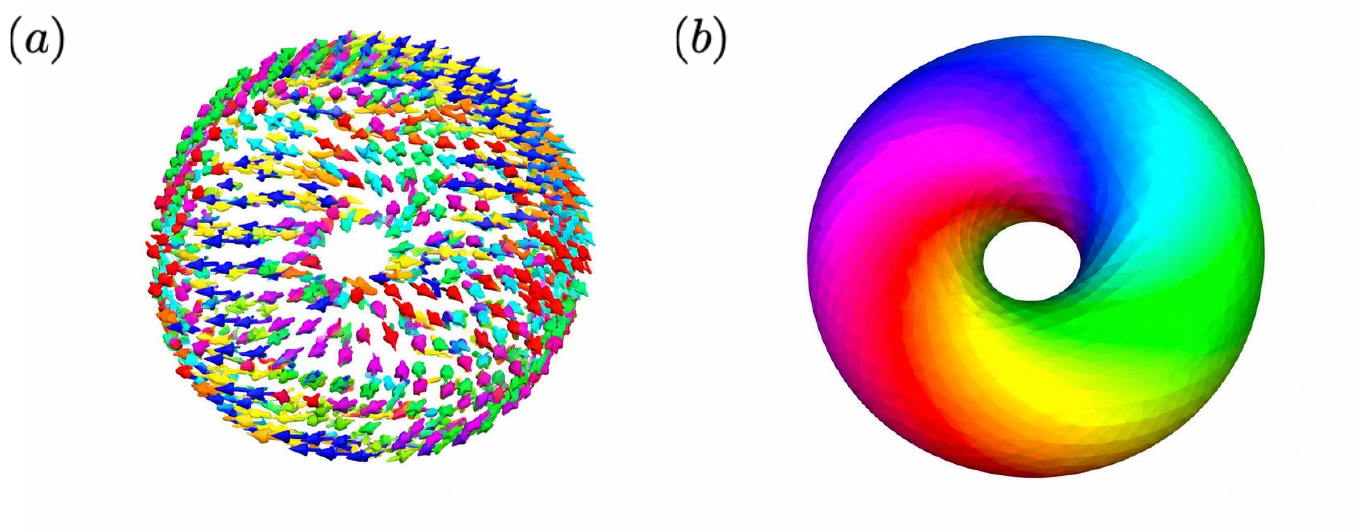}
\caption{Numerically stabilized hopfion in the model Eq. \eqref{Micro_S2}, representing the (a) spin field and (b) Néel field, respectively.}
\label{Figure4: hopfion-solution}
\end{figure}

\section{\NoCaseChange{Microscopic derivation of the tensor rotation-field energy}}\label{sec:sm-tensor-microscopic}
We derive the tensor functional of main-text Eq.~\eqref{eq:continuum-energy}, retaining the spatial and spin-frame structure of its coefficients. Following the main text, the reference spin directions are $\mathbf e_\mu$ and the rigid-frame parametrization is
\begin{equation}
\mathbf S(\mathbf R_{n,\mu}) =\mathcal R(\mathbf R_n)\mathbf e_\mu,
\qquad \mu=0,1,2,3, \qquad \mathcal R^T\mathcal R=\mathbbm{1},
\qquad \det\mathcal R=1.
\label{eq:sm-frame}
\end{equation}
It retains common rotations of a fixed noncollinear reference state, not independent distortions of the relative sublattice angles. Rotation fields of this type provide the order-parameter description of noncollinear magnets~\cite{zarzuela2019hydrodynamics}.

Let $v_c$ be the volume of the cell used in Eq.~\eqref{eq:sm-frame}. A directed bond $b=(\mu,\nu,\boldsymbol\delta_b)$ connects a spin of sublattice $\mu$ in cell $\mathbf R_n$ to a spin of sublattice $\nu$ in cell $\mathbf R_n+\boldsymbol\delta_b$. Since the frame is evaluated at cell positions, $\boldsymbol\delta_b$ is the displacement between those positions. A site-centered interpolation would instead require site displacements consistently throughout the derivation. The bond set $\mathscr B$ contains both orientations of each bond, with
\begin{equation}
\bar b=(\nu,\mu,-\boldsymbol\delta_b),
\qquad J_{\bar b}=J_b,
\qquad \mathbf D_{\bar b}=-\mathbf D_b.
\label{eq:sm-bonds}
\end{equation}
Thus an unordered bond sum is represented by $\tfrac12\sum_n\sum_{b\in\mathscr B}$ and $\sum_n\to v_c^{-1}\int d^3r$. The notation $J_b$ labels the exchange on each retained bond. The expansion retains even exchange gradients through fourth order and DMI through first order. 

\paragraph{Exchange moments:} Substituting Eq.~\eqref{eq:sm-frame} into the exchange energy gives
\begin{equation}
E_J=\frac{1}{2v_c}\int d^3r\sum_{b\in\mathscr B}J_b\,
\mathbf e_\mu^T\mathcal R^T(\mathbf r)
\mathcal R(\mathbf r+\boldsymbol\delta_b)\mathbf e_\nu.
\label{eq:sm-exchange-start}
\end{equation}
For a field varying slowly on the bond scale,
\begin{equation}
\mathcal R(\mathbf r+\boldsymbol\delta_b)
=\sum_{m=0}^{4}\frac{\delta_b^{i_1}\cdots\delta_b^{i_m}}{m!}
\partial_{i_1}\cdots\partial_{i_m}\mathcal R(\mathbf r)
+\text{higher-gradient terms},
\label{eq:sm-taylor}
\end{equation}
where the $m=0$ term is $\mathcal R$. This expansion makes no
small-angle approximation. The zeroth-order density is independent of $\mathcal R$ and is subtracted. At second order, one integration by parts yields
\begin{align}
E_J^{(2)}
&=\frac{1}{4v_c}\int d^3r\sum_b
J_b\delta_b^i\delta_b^j\,
\mathbf e_\mu^T\mathcal R^T\partial_i\partial_j\mathcal R
\mathbf e_\nu \nonumber\\
&=-\frac{1}{4v_c}\int d^3r\sum_b
J_b\delta_b^i\delta_b^j\,
\mathbf e_\mu^T(\partial_i\mathcal R)^T
\partial_j\mathcal R\mathbf e_\nu =\int d^3r\,\operatorname{Tr}\!\left[
G^{ij}(\partial_i\mathcal R)^T\partial_j\mathcal R\right],
\label{eq:sm-J2}
\end{align}
with
\begin{equation}
G^{ij}=-\frac{1}{4v_c}\sum_b
J_b\delta_b^i\delta_b^j\,\mathbf e_\nu\mathbf e_\mu^T.
\label{eq:sm-G}
\end{equation}
Here we used
$\mathbf u^TM\mathbf v=\operatorname{Tr}(\mathbf v\mathbf u^TM)$. At fourth order, two integrations by parts give
\begin{align}
E_J^{(4)}
&=\frac{1}{48v_c}\int d^3r\sum_b
J_b\delta_b^i\delta_b^j\delta_b^k\delta_b^l\,
\mathbf e_\mu^T\mathcal R^T
\partial_i\partial_j\partial_k\partial_l\mathcal R\mathbf e_\nu
\nonumber\\
&=\frac{1}{48v_c}\int d^3r\sum_b J_b\delta_b^i\delta_b^j\delta_b^k\delta_b^l\,
\mathbf e_\mu^T(\partial_i\partial_j\mathcal R)^T
\partial_k\partial_l\mathcal R\mathbf e_\nu=\int d^3r\,\operatorname{Tr}\!\left[
\mathsf W^{ijkl}(\partial_i\partial_j\mathcal R)^T
\partial_k\partial_l\mathcal R\right],
\label{eq:sm-J4}
\end{align}
where $\mathsf W^{ijkl}=\frac{1}{48v_c}\sum_b
J_b\delta_b^i\delta_b^j\delta_b^k\delta_b^l\,
\mathbf e_\nu\mathbf e_\mu^T.$

\paragraph{DMI contributions:} The DMI energy becomes
\begin{equation}
E_{\rm DM}=\frac{1}{2v_c}\int d^3r\sum_b
\mathbf D_b\cdot\left[
(\mathcal R\mathbf e_\mu)\times
\bigl(\mathcal R(\mathbf r+\boldsymbol\delta_b)\mathbf e_\nu\bigr)
\right].
\label{eq:sm-D-start}
\end{equation}
Define the DMI vector in the local frame,
$\widetilde{\mathbf D}_b(\mathcal R)=\mathcal R^T\mathbf D_b$. At first order, $\partial_i\mathcal R=\mathcal R\mathcal L_i$ gives
\begin{equation}
\mathcal E_{\rm DM}^{(1)}
=\frac{1}{2v_c}\sum_b\delta_b^i
\widetilde{\mathbf D}_b(\mathcal R)\cdot
\left[\mathbf e_\mu\times(\mathcal L_i\mathbf e_\nu)\right]
=\Lambda_i^a(\mathcal R)A_i^a,
\label{eq:sm-D1}
\end{equation}
where
\begin{equation}
\Lambda_i^a(\mathcal R)
=\frac{1}{2v_c}\sum_b\delta_b^i
\widetilde{\mathbf D}_b(\mathcal R)\cdot
\left[\mathbf e_\mu\times(\hat L_a\mathbf e_\nu)\right].
\label{eq:sm-Lambda}
\end{equation}
Choosing the antisymmetric representative $H^i(\mathcal R)=-\Lambda_i^a(\mathcal R)\hat L_a/2$ yields
\begin{equation}
\mathcal E_{\rm DM}^{(1)}
=\operatorname{Tr}[H^i(\mathcal R)\mathcal L_i].
\label{eq:sm-D1-trace}
\end{equation}
Only the antisymmetric part of $H^i$ contributes. 

\paragraph{Magnetic anisotropy:} The single-ion term follows directly from the rigid-frame ansatz:
\begin{equation}
E_{\rm an}=-\frac{K}{v_c}\int d^3r\sum_\mu (\mathbf e_\mu^T\mathcal R\mathbf e_\mu)^2 =K\int d^3r\,U(\mathcal R),
\end{equation}
with $\Pi_\mu=\mathbf e_\mu\mathbf e_\mu^T,$ and
\begin{equation}
U(\mathcal R)=-\frac{1}{v_c}\sum_\mu[\operatorname{Tr}(\Pi_\mu\mathcal R)]^2.
\label{eq:sm-U}
\end{equation}
The coefficient $K$ retains its microscopic meaning; the conversion to energy density is contained in $U$. 

Combining the retained contributions gives
\begin{align}
\mathcal E_{\rm tot}=\operatorname{Tr}\!\left[ G^{ij}(\partial_i\mathcal R)^T\partial_j\mathcal R +H^i(\mathcal R)\mathcal L_i\right]+\operatorname{Tr}\!\left[ \mathsf W^{ijkl}(\partial_i\partial_j\mathcal R)^T\partial_k\partial_l\mathcal R\right]+K U(\mathcal R).
\label{eq:sm-tensor-energy}
\end{align}
This is main-text Eq.~(4). 

\section{\NoCaseChange{Derrick--Hobart analysis}}
\label{sec:sm-tensor-derrick}
Assume homogeneous material coefficients and a smooth localized
reference texture $\mathcal R_0(\mathbf r)$ on $\mathbb R^3$, with
$\mathcal R_0\to\mathcal R_\infty$ and finite excess energy.
The Derrick--Hobart family~\cite{Derrick1964,Hobart1963} is $\mathcal R_\lambda(\mathbf r)=\mathcal R_0(\mathbf r/\lambda)$. $\mathbf y=\mathbf r/\lambda$, and $\lambda>0$. It preserves the rotation constraint and the fixed asymptotic state. The field-dependent coefficient satisfies $H^i(\mathcal R_\lambda(\mathbf r))=H^i(\mathcal R_0(\mathbf y))$. Define the four reference energies using the full tensors:
\begin{align}
C_J&=\int d^3y\,\operatorname{Tr}\!\left[
G^{ij}(\partial_i\mathcal R_0)^T\partial_j\mathcal R_0\right],
\label{eq:sm-CJ}\\
C_D&=\int d^3y\,\operatorname{Tr}\!\left[
H^i(\mathcal R_0)\mathcal L_{0i}\right],
\label{eq:sm-CD}\\
C_W&=\int d^3y\,\operatorname{Tr}\!\left[
\mathsf W^{ijkl}(\partial_i\partial_j\mathcal R_0)^T
\partial_k\partial_l\mathcal R_0\right],
\label{eq:sm-CW}\\
C_K&=\int d^3y\,K\Delta U(\mathcal R_0).
\label{eq:sm-CK}
\end{align}
Direct substitution in Eq.~\eqref{eq:sm-tensor-energy} yields
\begin{equation}
{E_{\rm exc}(\lambda)
=C_J\lambda+C_D\lambda^2+C_K\lambda^3+\frac{C_W}{\lambda}.}
\label{eq:sm-scaling}
\end{equation}
The exponents depend only on derivative order. 

\paragraph{Stability:} Differentiating Eq.~\eqref{eq:sm-scaling},
\begin{align}
E_{\rm exc}'(\lambda)
&=C_J+2C_D\lambda+3C_K\lambda^2-\frac{C_W}{\lambda^2},
\label{eq:sm-first-derivative}\\
E_{\rm exc}''(\lambda)
&=2C_D+6C_K\lambda+\frac{2C_W}{\lambda^3}.
\label{eq:sm-second-derivative}
\end{align}
Choosing a stationary reference texture with scale $\lambda=1$ gives $C_J+2C_D+3C_K-C_W=0.$ Strict quadratic stability along the dilation direction requires
\begin{equation}
E_{\rm exc}''(1)=2(C_D+3C_K+C_W)>0.
\label{eq:sm-scale-stability}
\end{equation}
For a given profile, $C_W>0$ makes the retained fourth-gradient energy diverge under uniform contraction.  In the exchange--frustration limit, $C_D=C_K=0$, positive $C_J$ and
$C_W$ yield
\begin{equation}
\lambda^{(0)}=\sqrt{\frac{C_W}{C_J}},
\qquad
E_{\rm exc}''(\lambda^{(0)})
=\frac{2C_W}{[\lambda^{(0)}]^3}>0.
\label{eq:sm-lambda0}
\end{equation}
The competition of second- and fourth-gradient contributions thus
sets a finite scale without DMI, within this fixed-profile family.
This is the scaling mechanism underlying the comparison with
Skyrme stabilization~\cite{Skyrme1961}.

Keep $G^{ij}$, $\mathsf W^{ijkl}$, and the reference profile fixed,
and take $C_D,C_K=O(\varepsilon)$ relative to the exchange--frustration
energies. The dimensionless smallness conditions are
\begin{equation}
\left|\frac{2C_D\lambda^{(0)}}{C_J}\right|\ll1,
\qquad
\left|\frac{3C_K[\lambda^{(0)}]^2}{C_J}\right|\ll1.
\label{eq:sm-smallness}
\end{equation}
Set $\lambda_\star=\lambda^{(0)}+\delta\lambda$ with
$\delta\lambda=O(\varepsilon)$. Expanding
$E_{\rm exc}'(\lambda_\star)=0$ gives
\begin{equation}
2C_D\lambda^{(0)}+3C_K[\lambda^{(0)}]^2
+\frac{2C_W}{[\lambda^{(0)}]^3}\delta\lambda
=O(\varepsilon^2),
\end{equation}
where $C_J-C_W/[\lambda^{(0)}]^2=0$ has been used. Hence
\begin{align}
\delta\lambda=-\frac{[\lambda^{(0)}]^3}{2C_W}
\left[2C_D\lambda^{(0)}+3C_K[\lambda^{(0)}]^2\right]
+O(\varepsilon^2)=-\frac{C_D C_W}{C_J^2}
-\frac{3C_K C_W^{3/2}}{2C_J^{5/2}}
+O(\varepsilon^2).
\label{eq:sm-deltalambda}
\end{align}
If the reference profile is normalized so that $\lambda_\star=1$,
\begin{equation}
\sqrt{\frac{C_W}{C_J}}
-\frac{C_D C_W}{C_J^2}
-\frac{3C_K C_W^{3/2}}{2C_J^{5/2}}
=1+O(\varepsilon^2).
\label{eq:sm-unit-scale}
\end{equation}
This is the perturbative stationarity relation, describing the scale shift at fixed shape.

\section{\NoCaseChange{Quaternion representation}}
\label{sec:sm-tensor-quaternion}
Write a unit quaternion as $q=(q_0,\mathbf q)$, with conjugation
$q^*=(q_0,-\mathbf q)$ and Hamilton product
\begin{equation}
(p_0,\mathbf p)(q_0,\mathbf q)
=(p_0q_0-\mathbf p\cdot\mathbf q,\,
p_0\mathbf q+q_0\mathbf p+\mathbf p\times\mathbf q).
\label{eq:sm-Hamilton-product}
\end{equation}
Thus $q^*q=q_0^2+|\mathbf q|^2=1$. We fix the rotation convention by
$\mathcal R(q)\mathbf v=q\mathbf vq^*$ for a vector identified with
the purely imaginary quaternion $(0,\mathbf v)$.
Equivalently~\cite{zarzuela2019hydrodynamics},
\begin{equation}
\mathcal R_{ab}(q)
=(q_0^2-|\mathbf q|^2)\delta_{ab}
+2q_aq_b-2q_0\epsilon_{abc}q_c.
\label{eq:sm-Rq}
\end{equation}
Both $q$ and $-q$ represent the same rotation. With the generators,
$\hat L(\mathbf u)\mathbf v=-\mathbf u\times\mathbf v$,
where $\hat L(\mathbf u)=u^a\hat L_a$. The Shankar ansatz
$\mathcal R_S=\exp[f(r)\hat{\mathbf r}\cdot\hat{\mathbf L}]$
therefore has the lift
$q_S=(\cos[f(r)/2],-\hat{\mathbf r}\sin[f(r)/2])$.
This fixes the signs below without changing the main-text ansatz.

Introduce the combinations
\begin{align}
\mathbf a_i
&=\operatorname{Im}(q^*\partial_iq)
=q_0\partial_i\mathbf q-\mathbf q\,\partial_iq_0
-\mathbf q\times\partial_i\mathbf q,
\label{eq:sm-ai}\\
\mathbf b_{ij}
&=\operatorname{Im}(q^*\partial_i\partial_jq)
=q_0\partial_i\partial_j\mathbf q
-\mathbf q\,\partial_i\partial_jq_0
-\mathbf q\times\partial_i\partial_j\mathbf q.
\label{eq:sm-bij}
\end{align}
They are determined by $q$ and its derivatives, not independent
fields. Unit norm implies
\begin{align}
\operatorname{Re}(q^*\partial_iq)&=0,
&g_{ij}&\equiv\operatorname{Re}[(\partial_iq)^*\partial_jq]
=\mathbf a_i\cdot\mathbf a_j,
\nonumber\\
\operatorname{Re}(q^*\partial_i\partial_jq)&=-g_{ij},
&\partial_iq&=q(0,\mathbf a_i).
\label{eq:sm-unit-identities}
\end{align}
For a fixed vector $\mathbf v$, differentiating $q\mathbf vq^*$
and rotating back to the reference frame gives
\begin{equation}
\mathcal R^T\partial_i\mathcal R\,\mathbf v
=(0,\mathbf a_i)\mathbf v-\mathbf v(0,\mathbf a_i)
=2\mathbf a_i\times\mathbf v.
\end{equation}
It follows that $\mathcal L_i=-2a_i^a\hat L_a,$ and $ A_i^a=-2a_i^a.$

\paragraph{Tensor exchange energy:} The generator identity $\hat L(\mathbf u)\hat L(\mathbf v)
=\mathbf v\mathbf u^T-(\mathbf u\cdot\mathbf v)\mathbbm{1}$ gives $(\partial_i\mathcal R)^T\partial_j\mathcal R =\mathcal L_i^T\mathcal L_j
=4\left[g_{ij}\mathbbm{1}-\mathbf a_j\mathbf a_i^T\right].$  Consequently,
\begin{align}
\mathcal E_J^{(2)}[q]=4\left[\operatorname{Tr}(G^{ij})
(\mathbf a_i\cdot\mathbf a_j)
-\operatorname{Tr}(G^{ij}\mathbf a_j\mathbf a_i^T)\right]=4\mathbf a_i^T
\left[\operatorname{Tr}(G^{ij})\mathbbm{1}-G^{ij}\right]
\mathbf a_j.
\label{eq:sm-Jq}
\end{align}

For the fourth-gradient energy we need each second derivative of
$\mathcal R$, rather than only its Laplacian. Differentiating
$\partial_j\mathcal R=\mathcal R\mathcal L_j$ yields $\mathcal R^T\partial_i\partial_j\mathcal R
=\partial_i\mathcal L_j+\mathcal L_i\mathcal L_j.$ On the quaternion side, $\partial_iq^*=-q^*(\partial_iq)q^*$ implies
\begin{equation}
\partial_i(q^*\partial_jq)
=-(q^*\partial_iq)(q^*\partial_jq)
+q^*\partial_i\partial_jq.
\end{equation}
Taking imaginary parts gives $\partial_i\mathbf a_j
=\mathbf b_{ij}-\mathbf a_i\times\mathbf a_j.$ Then,
\begin{align}
\mathcal B_{ij}[q]
\equiv\mathcal R^T\partial_i\partial_j\mathcal R
={}&-2b_{ij}^a\hat L_a
+2(\mathbf a_i\mathbf a_j^T+\mathbf a_j\mathbf a_i^T)
-4g_{ij}\mathbbm{1}.
\label{eq:sm-Bmatrix}
\end{align}
The cancellation of the antisymmetric part of the dyadic products
uses
$\hat L(\mathbf a_i\times\mathbf a_j)
=\mathbf a_i\mathbf a_j^T-\mathbf a_j\mathbf a_i^T$.
In components,
\begin{equation}
(\mathcal B_{ij})_{ab}
=-2\epsilon_{abc}b_{ij}^c
+2a_i^aa_j^b+2a_j^aa_i^b-4g_{ij}\delta_{ab}.
\label{eq:sm-Bcomponents}
\end{equation}
The resulting identity is: $\partial_i\partial_j\mathcal R=\mathcal R\mathcal B_{ij}$ and $(\partial_i\partial_j\mathcal R)^T\partial_k\partial_l\mathcal R
=\mathcal B_{ij}^T\mathcal B_{kl}.$ Hence
\begin{equation}
\mathcal E_J^{(4)}[q]
=\operatorname{Tr}\!\left[
\mathsf W^{ijkl}\mathcal B_{ij}^T\mathcal B_{kl}\right].
\label{eq:sm-Wq}
\end{equation}
The contraction is explicitly
$\sum_{ijkl}\sum_{abc}(\mathsf W^{ijkl})_{ab}
(\mathcal B_{ij})_{cb}(\mathcal B_{kl})_{ca}$.

\paragraph{DMI and anisotropy:} Decompose the antisymmetric part of the original DMI matrix as
\begin{equation}
H^i_{\rm as}(\mathcal R(q))=h_i^a(q)\hat L_a,
\qquad
h_i^a(q)=-\frac12\operatorname{Tr}
[H^i(\mathcal R(q))\hat L_a],
\label{eq:sm-hdef}
\end{equation}
and since $h_i^a=-\Lambda_i^a/2$, we find
\begin{equation}
\mathcal E_{\rm DM}^{(1)}[q]
=\operatorname{Tr}[H^i\mathcal L_i]
=4\mathbf h_i(q)\cdot\mathbf a_i.
\label{eq:sm-Dq}
\end{equation}
The same microscopic coefficients can be written entirely in quaternion components. From Eq.~\eqref{eq:sm-Rq},
\begin{equation}
\widetilde{\mathbf D}_b(q)
=(q_0^2-|\mathbf q|^2)\mathbf D_b
+2\mathbf q(\mathbf q\cdot\mathbf D_b)
-2q_0\,\mathbf q\times\mathbf D_b.
\label{eq:sm-Dtildeq}
\end{equation}
Using $\mathcal L_i\mathbf e_\nu=2\mathbf a_i\times\mathbf e_\nu$ Eq.~\eqref{eq:sm-D1} yields
\begin{equation}
\mathbf h_i(q)=\frac{1}{4v_c}\sum_b\delta_b^i
\left[(\mathbf e_\mu\cdot\mathbf e_\nu)
\widetilde{\mathbf D}_b(q)
-\bigl(\widetilde{\mathbf D}_b(q)\cdot\mathbf e_\nu\bigr)
\mathbf e_\mu\right].
\label{eq:sm-hbonds}
\end{equation}
For the main-text anisotropy projectors,
\begin{equation}
\operatorname{Tr}(\Pi_\mu\mathcal R(q))
=\mathbf e_\mu^T\mathcal R(q)\mathbf e_\mu
=q_0^2-|\mathbf q|^2+2(\mathbf q\cdot\mathbf e_\mu)^2,
\end{equation}
and therefore
\begin{equation}
U(q)=-\frac{1}{v_c}\sum_\mu
\left[q_0^2-|\mathbf q|^2+2(\mathbf q\cdot\mathbf e_\mu)^2\right]^2.
\label{eq:sm-Uq}
\end{equation}

\newpage

\section{\NoCaseChange{MuMax3 script for the frustrated antiferromagnet}\label{mumax3_script}}
\begin{lstlisting}
OutputFormat = OVF2_BINARY

A := -4.1e-12; Aex = A; 				      //Exchnage stiffness
n := 64; dx := 1e-8; L := dx*n; 		  //System discretization and size
SetMesh(n, n, n, dx, dx, dx, 1, 1, 1) //cube geometry
Ms := -2.0*A/(0.205*dx*dx) 				    //for hopfion
// Ms := -2.0*A/(0.2*dx*dx) 			    //for Shankar
Msat = Ms;								            //Msat is internally used in mumax.
MinimizerStop = 1e-5
EnableDemag = false

//**** Implementation of AFM frustration via custom fields
prefA1 	  := Const(-0.05)

left1 	  := Mul(Add(Mul(Const(-1), m),Shifted(m,2,0,0)), Shifted(Const(1),2,0,0))
left2 	  := Mul(Add(Mul(Const(-1),m),Shifted(m,2-n,0,0)),Shifted(Const(1),2-n,0,0))
left  	  := Add(left1, left2)

right1 	  := Mul(Add(Mul(Const(-1),m),Shifted(m,-2,0,0)),Shifted(Const(1),-2,0,0))
right2 	  := Mul(Add(Mul(Const(-1),m),Shifted(m,n-2,0,0)),Shifted(Const(1),n-2,0,0))
right  	  := Add(right1, right2)

backward1 := Mul(Add(Mul(Const(-1),m),Shifted(m,0,2,0)), Shifted(Const(1),0,2,0))
backward2 := Mul(Add(Mul(Const(-1),m),Shifted(m,0,2-n,0)),Shifted(Const(1),0,2-n,0))
backward  := Add(backward1, backward2)

forward1  := Mul(Add(Mul(Const(-1),m),Shifted(m,0,-2,0)),Shifted(Const(1),0,-2,0))
forward2  := Mul(Add(Mul(Const(-1),m),Shifted(m,0,n-2,0)),Shifted(Const(1),0,n-2,0))
forward   := Add(forward1, forward2)

down1 	  := Mul(Add(Mul(Const(-1),m),Shifted(m,0,0,2)), Shifted(Const(1),0,0,2))
down2     := Mul(Add(Mul(Const(-1),m),Shifted(m,0,0,2-n)),Shifted(Const(1),0,0,2-n))
down  	  := Add(down1, down2)

up1 	   := Mul(Add(Mul(Const(-1),m),Shifted(m,0,0,-2)),Shifted(Const(1),0,0,-2))
up2 	   := Mul(Add(Mul(Const(-1),m),Shifted(m,0,0,n-2)),Shifted(Const(1),0,0,n-2))
up  	   := Add(down1, down2)

XA 		  := Add(Mul(prefA1, left), 	Mul(prefA1, right))
YA 		  := Add(Mul(prefA1, forward), 	Mul(prefA1, backward))
ZA  	  := Add(Mul(prefA1, down), 	Mul(prefA1, up))

BcA 	  := Add(XA, Add(YA, ZA))
AddFieldTerm(BcA)
addEdensTerm(Mul(Const(-0.5), Dot(BcA, M_full)))

//****Initial state
m.LoadFile("AFM_hopf.ovf") //hopfion
// m.LoadFile("AFM_Shan.ovf") //Shankar monopole

relax()
save(m)
minimize()
save(m)
/*
saving Neel vector field
only implemented in mumax3-gneb
https://kuchkin.github.io/download.html 
*/
// saveas(Flipped(m), "nn.ovf") 

\end{lstlisting}

\newpage

\section{\NoCaseChange{Atomistic spin dynamics simulations and spectral analysis}}\label{sec:sm-dynamics}

The low amplitude dynamics discussed in the main text was obtained from atomistic spin dynamics simulations performed with Magnoom~\cite{savchenko2022chiral}. We consider the same frustrated Heisenberg antiferromagnet used to stabilize the textures, on a $64^3$ lattice with periodic boundary conditions. For the Shankar skyrmion we use $J_1=0.20$ and $J_4=0.05$, in the exchange-sign convention of the main text, while the hopfion is simulated with $J_1=0.205$ and $J_4=0.05$. The dynamics is integrated with the SIB scheme, using a time step $\Delta t=0.1$ and a Gilbert damping $\alpha=10^{-4}$.

Starting from the relaxed textures, we perturb the system with a single rectangular easy-axis anisotropy pulse,
\begin{equation}
K_z \left( t \right)=K^0_z \left[ \Theta \left( t \right)-\Theta \left( t-\tau_{\mathrm{p}} \right) \right], 
\label{eq:SM_pulse}
\end{equation}
being $K^0_z=10^{-5}$ the dimensionless anisotropy pulse amplitude and $\tau_{\mathrm{p}}$ the dimensionless pulse duration, which in this case lasts $\tau_{\mathrm{p}}=1$, corresponding to 10 integration steps. Its role is not to resonantly drive a selected excitation, but to weakly perturb the relaxed texture over a short time interval, which contains a broad range of frequencies and allows several dynamical excitations to be probed from the subsequent free evolution within a single simulation. We use an anisotropy pulse rather than a magnetic field pulse because it perturbs the local energy landscape of the non-collinear texture without directly imposing a Zeeman torque or a preferred direction of motion. It therefore provides a simple way of coupling to internal deformations of the texture, while still allowing extended spin wave excitations to be present in the broadband response. The small amplitude was chosen to preserve the texture and probe its low amplitude response rather than induce a large-scale reconfiguration. We therefore use the pulse as a spectroscopic perturbation and analyze the dynamics only after it has been removed.

For comparison, we also evolved the same relaxed configurations without applying the anisotropy pulse. These undriven trajectories provide a reference for the weak residual dynamics of the numerically relaxed state. The resonances discussed below are identified from the additional response induced by the pulse rather than from spectral features already present in this background evolution.

\subsection{Spectral response}

We first characterize the post-pulse dynamics from the spatially averaged N\'eel vector,
\begin{equation}
\overline{\mathbf{n}} \left( t \right)= \frac{1}{N} \sum_{\mathbf{r}} \mathbf{n} \left( \mathbf{r}, t \right),
\label{eq:SM_average}
\end{equation}
where the sum runs over all $N=64^3$ lattice sites. The three Cartesian components of $\bar{\mathbf n}(t)$ are detrended to remove the slowly varying background associated with residual relaxation and multiplied by a Hann window before Fourier transformation~\cite{harris1978use}. We denote the resulting Fourier amplitude of the $\mu$-th component by
$\tilde{n}_{\mu}(f)$, where $\mu=x,y,z$ and $f$ denotes the frequency. We then combine the three components into the scalar spectral response,
\begin{equation}
A \left( f \right)=\sqrt{\left| \widetilde{n}_x \left( f \right) \right|^2+\left| \widetilde{n}_y \left( f \right) \right|^2+\left| \widetilde{n}_z \left( f \right) \right|^2},
\label{eq:SM_spectrum}
\end{equation} 
This definition combines the response of the three Cartesian components and avoids assigning a resonance on the basis of a particular projection of the dynamics.

The finite duration of the simulated trajectories, $2.5 \times 10^4$, sets the frequency resolution to $\Delta f=1/T \simeq 4 \times 10^{-5}$. For plotting, we zero-pad the time traces before computing the Fourier transform. This adds intermediate points to the spectrum, making the peaks appear smoother, but does not add information or improve the frequency resolution.

As a check of the identified resonances, we repeated the spectral analysis using different time windows and detrending procedures. We also compared the Fourier spectrum obtained from the complete trajectory with a Welch spectrum, in which the trajectory is divided into overlapping segments and the resulting spectra are averaged~\cite{welch1967use}. The frequencies of the three resonances discussed in the main text remain stable under these different analyses. The full-record and Welch spectra give, respectively,
\begin{align}
f_{\mathrm{S1}} &= 1.1800\times10^{-3},
& f_{\mathrm{S1}}^{\mathrm{Welch}} &= 1.1826 \times 10^{-3},\\
f_{\mathrm{H1}} &= 1.6750 \times 10^{-3},
& f_{\mathrm{H1}}^{\mathrm{Welch}} &=1.6785 \times 10^{-3},\\
f_{\mathrm{S2}} &= 2.6100 \times 10^{-3},
& f_{\mathrm{S2}}^{\mathrm{Welch}} &=2.6131 \times 10^{-3}.
\end{align}

The frequencies above are expressed in the reduced time units used by
Magnoom~\cite{savchenko2022chiral}. The dimensionless simulation time $t_{\mathrm{M}}$ is related
to physical time by
\begin{equation}
t_{\mathrm{M}}
=
t_{\mathrm{phys}}\,\gamma\frac{E_0}{\mu_s},
\label{eq:time_conversion}
\end{equation}
where $\gamma$ is the gyromagnetic ratio, $E_0$ is the physical energy
corresponding to one reduced energy unit, and $\mu_s$ is the atomic
magnetic moment. Consequently, the corresponding physical frequency is
\begin{equation}
f_{\mathrm{phys}}
=
f_{\mathrm{M}}\,\gamma\frac{E_0}{\mu_s}.
\label{eq:frequency_conversion}
\end{equation}
A conversion to an absolute frequency therefore requires a physical
parametrization of $E_0$ and $\mu_s$. As an illustrative example, taking
$E_0=1~\mathrm{meV}$, $\mu_s=1~\mu_{\mathrm{B}}$, and
$\gamma=1.7609\times10^{11}~\mathrm{s^{-1}T^{-1}}$ gives
$f_{\mathrm{S1}}\simeq3.59~\mathrm{GHz}$,
$f_{\mathrm{H1}}\simeq5.09~\mathrm{GHz}$, and
$f_{\mathrm{S2}}\simeq7.94~\mathrm{GHz}$.
These values are illustrative rather than material-specific and scale
linearly with $E_0/\mu_s$.

\begin{figure}[t]
\centering
\includegraphics[width=\linewidth]{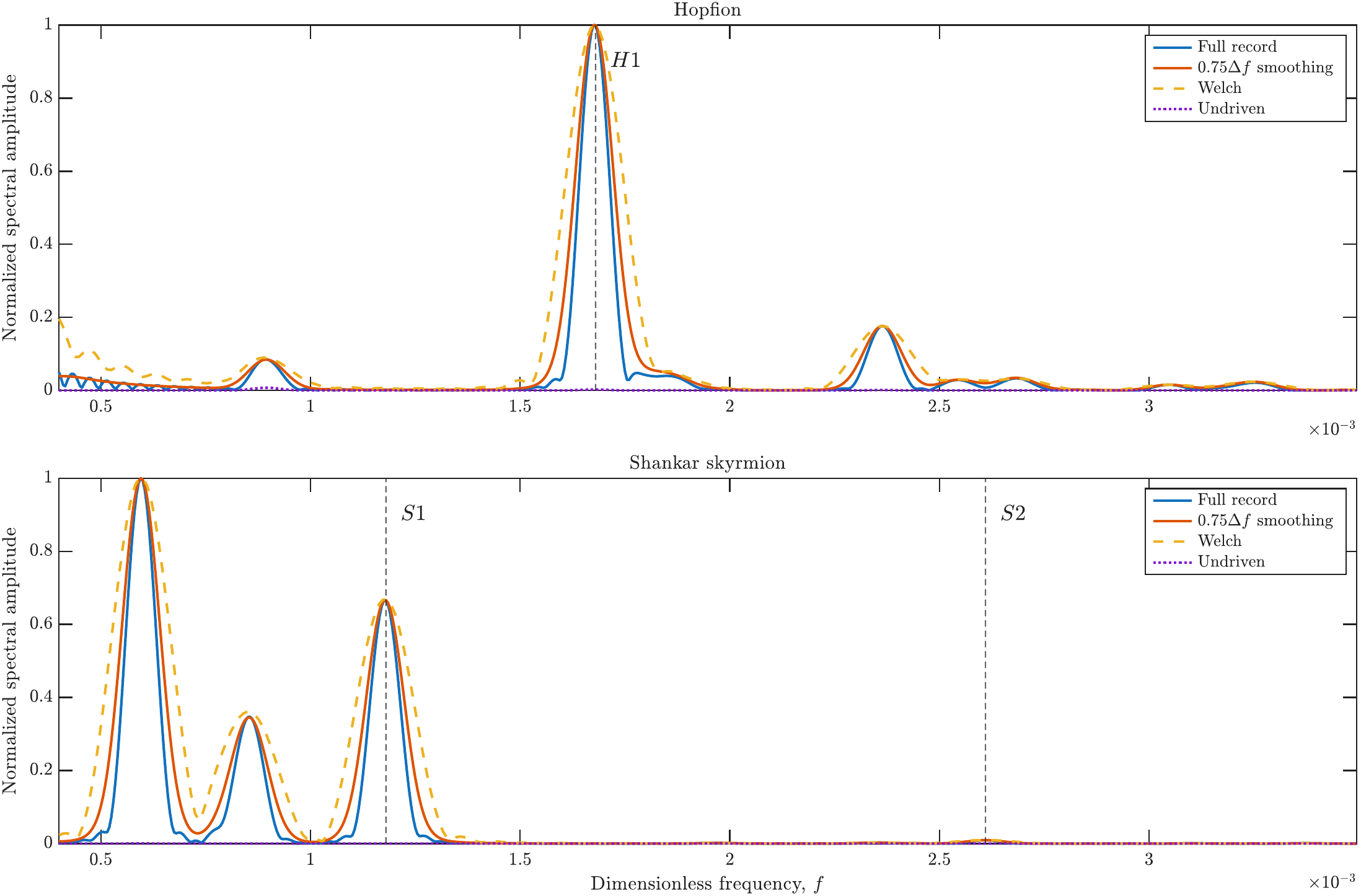}
\caption{Robustness of the spectral analysis for the hopfion and Shankar skyrmion. The full-record Fourier spectrum is compared with the spectrum after the weak Gaussian smoothing used in Fig.~\ref{FFT}(a), with
$\sigma=0.75 \Delta f$, the Welch spectral estimate, and the corresponding undriven reference. All curves are normalized to the maximum of the driven full-record spectrum for each texture, so that the magnitude of the residual undriven response is preserved. The positions of the texture-associated resonances H1, S1, and S2 remain stable under the different spectral analyses.}
\label{fig:SM_spectral_checks}
\end{figure}

The spectra displayed in Fig.~\ref{FFT}(a) are lightly Gaussian-smoothed for visual clarity, using $\sigma=0.75 \Delta f \simeq 3 \times 10^{-5}$. We verified the evolution of the spectral profile for $\sigma/\Delta f=0$, $0.5$, $0.75$, $1.00$, and $1.25$. The smoothing suppresses small oscillations associated with the finite time window without appreciably shifting the resonances retained in the analysis. Importantly, smoothing is used only for the graphical representation in Fig.~\ref{FFT}(a): the resonance frequencies are determined from the unsmoothed analysis and checked independently against the Welch spectrum. The hopfion and Shankar spectra in Fig.~\ref{FFT}(a) are also normalized independently to their respective maxima. Their relative heights should therefore not be interpreted as a comparison of the absolute response amplitudes of the two textures. The robustness of the identified resonances against these spectral processing choices, together with the corresponding undriven references, is summarized in Fig.~\ref{fig:SM_spectral_checks}.

\subsection{Spatial structures of the resonances}\label{sec:sm-spatial}

The presence of a peak in the spatially averaged spectrum is not by itself sufficient to identify an internal excitation of the texture. Extended spin waves, residual relaxation, and approximately rigid collective motion can all contribute to the Fourier spectrum. We therefore complement the spectral analysis with a spatially resolved harmonic decomposition of the N\'eel vector dynamics~\cite{yang2021intrinsic,bo2021spin}. For each selected frequency, the spatial response is extracted directly from the driven trajectory. At every lattice site, the time-dependent N\'eel vector is fitted simultaneously to a quadratic slow background and to cosine and sine components at all selected frequencies,
\begin{equation}
\mathbf{n}(\mathbf{r},t)
=
\mathbf{c}_0(\mathbf{r})
+\mathbf{c}_1(\mathbf{r})\tau
+\mathbf{c}_2(\mathbf{r})\tau^2
+\sum_k
\left[
\mathbf{a}_k(\mathbf{r})\cos(2\pi f_k t)
+\mathbf{b}_k(\mathbf{r})\sin(2\pi f_k t)
\right],
\label{eq:SM_harmonic_fit}
\end{equation}
where $\tau=2 \left( t-t_{\min} \right)/\left( t_{\max}-t_{\min} \right)-1$ is the rescaled time variable used for the slowly varying background. The complex harmonic response at frequency $f_k$ is then defined as
\begin{equation}
\delta\mathbf{n}(\mathbf{r};f_k)
=
\mathbf{a}_k(\mathbf{r})
-\mathrm{i}\mathbf{b}_k(\mathbf{r}).
\label{eq:SM_complex}
\end{equation}
Its magnitude gives the local oscillation amplitude, while its phase retains the relative phase of the motion across the texture. This allows us to distinguish resonances that may have similar spatial amplitude profiles but correspond to different vectorial dynamics.

Part of this response can arise from a nearly rigid rotation of the whole
texture. To separate this contribution from an internal deformation, at each
frequency we determine the complex infinitesimal rotation vector
$\boldsymbol{\Theta}(f)$ by a least-squares fit over all lattice sites.
Specifically, $\boldsymbol{\Theta}(f)$ is chosen to minimize
\begin{equation}
\sum_{\mathbf r}
\left|
\delta\mathbf n(\mathbf r;f)
-
\boldsymbol{\Theta}(f)\times\mathbf n_0(\mathbf r)
\right|^2 .
\label{eq:rotation_fit}
\end{equation}
Here, $\mathbf n_0(\mathbf r)$ is the reference texture obtained from the
constant component of the simultaneous temporal fit at $\tau=0$, normalized
site by site. For this reference texture, the corresponding best-fit
rigid-rotation component is
\begin{equation}
\delta\mathbf n_{\rm rot}(\mathbf r;f)
=
\boldsymbol{\Theta}(f)\times\mathbf n_0(\mathbf r),
\label{eq:rotation_component}
\end{equation}
and we define the residual non-rigid response as
\begin{equation}
\delta\mathbf n_{\rm int}(\mathbf r;f)
=
\delta\mathbf n(\mathbf r;f)
-
\delta\mathbf n_{\rm rot}(\mathbf r;f).
\label{eq:internal_component}
\end{equation}
Here, the subscript ``int'' denotes the response remaining after removal of
the best-fit global rotation; by itself, it does not imply that the response
is a spatially localized internal mode of the texture.

The quantity plotted in Fig.~\ref{FFT}(b--d) is the magnitude of this non-rigid response,
\begin{equation}
A_{\mathrm{int}} \left( \mathbf{r}; f \right)=\left| \delta \mathbf{n}_{\mathrm{int}} \left( \mathbf{r}; f \right) \right|.
\label{eq:SM_internal_amplitude}
\end{equation}
Each selected-frequency response is normalized independently to its maximum amplitude, allowing their spatial profiles to be compared on the same scale. The maps therefore show the spatial structure of the internal response rather than its absolute strength. In particular, although both S1 and S2 reach unity after normalization, the absolute response of S2 is considerably weaker than that of S1.

To quantify the relative weights of the rigid-rotation and non-rigid contributions, we use their squared spatial norms relative to that of the complete harmonic response,
\begin{equation}
\eta_{\rm rot}(f)
=
\frac{
\sum_{\mathbf r}
|\delta\mathbf n_{\rm rot}(\mathbf r;f)|^2
}{
\sum_{\mathbf r}
|\delta\mathbf n(\mathbf r;f)|^2
},
\qquad
\eta_{\rm int}(f)
=
\frac{
\sum_{\mathbf r}
|\delta\mathbf n_{\rm int}(\mathbf r;f)|^2
}{
\sum_{\mathbf r}
|\delta\mathbf n(\mathbf r;f)|^2
}.
\label{eq:rigid_nonrigid_weights}
\end{equation}
Because $\delta\mathbf n_{\rm rot}$ is obtained by the least-squares projection defined above, the fitted rigid component and the residual are orthogonal under the corresponding spatial inner product, so that $\eta_{\rm rot}+\eta_{\rm int}=1$.

The maps in Fig.~\ref{FFT}(b--d) are central sections of the original $64^3$ response field. The H1 hopfion mode is shown in the central $xz$ section, while the S1 and S2 Shankar modes are shown in the central $xy$ section. The color interpolation used in plotting these maps only interpolates between the original lattice values and is not part of the spatial analysis.

\subsection{Identification of the texture-associated resonances}\label{sec:sm-mode-identification}

We distinguish texture-associated resonances from other spectral features using complementary spectral and spatial diagnostics: robustness of the resonance frequency against the spectral-analysis procedure, subtraction of the best-fit global rigid rotation, and spatial association of the remaining response with the equilibrium texture. This distinction is particularly
clear for the Shankar texture. Besides S1 and S2, its spectrum contains pronounced structures around $f \simeq 0.60 \times 10^{-3}$ and $f \simeq 0.86 \times 10^{-3}$. Their existence is robust in the Fourier spectrum, but their spatial response is extended over the simulation cell rather than concentrated around the Shankar texture. We therefore do not assign these peaks to internal Shankar modes.

\begin{figure}[t]
\centering
\includegraphics[width=\linewidth]{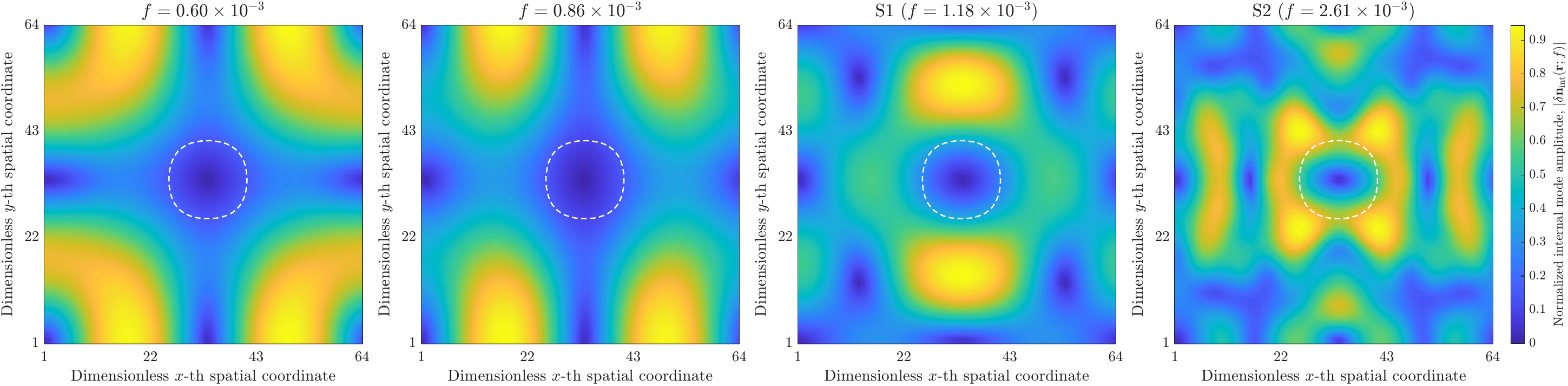}
\caption{Spatial classification of the Shankar skyrmion response. Normalized internal response amplitude, $|\delta\mathbf{n}_{\mathrm{int}}(\mathbf{r};f)|$, in the central $xy$ plane for the low-frequency features at $f \simeq 0.60 \times 10^{-3}$ and $0.86 \times 10^{-3}$ and for the texture-associated resonances S1 and S2. The same white dashed $n_0^z=0$ contour of the reference Shankar texture is shown in all panels as a spatial guide. The two low-frequency responses are broadly distributed throughout the simulation cell, whereas S1 and S2 exhibit spatial structures centered on and associated with the Shankar texture. Each map is normalized independently to its maximum. The comparison shows that a robust peak in the spatially averaged spectrum is not, by itself, sufficient to identify an internal texture resonance.}
\label{fig:SM_spatial_modes}
\end{figure}

For a simple quantitative measure of localization, we compare the average internal response amplitude in a central region of the three-dimensional simulation cell with that in an outer region. Here, $r$ denotes the distance from the center of the $64^3$ simulation cell, in lattice units. Taking $r\leq16$ for the central region and $r\geq26$ for the outer region gives central-to-outer amplitude ratios of approximately $0.32$ and $0.43$ for the low-frequency Shankar features. By contrast, the corresponding ratios are $2.13$ for S1 and $2.06$ for S2. The latter responses therefore exhibit a substantially stronger central concentration than the low-frequency features, as illustrated by the central cross-sections in Fig.~\ref{fig:SM_spatial_modes}.

The subtraction of the best-fit global rotation provides a complementary criterion. Using the squared-norm decomposition defined above, the non-rigid
fractions are $\eta_{\rm{int}} \simeq 99.98\%$ for S1 and
$\eta_{\rm{int}} \simeq 98.7\%$ for S2, corresponding to rigid-rotation fractions of only approximately $0.02\%$ and $1.3\%$, respectively. The dominant hopfion resonance H1 similarly retains approximately $94.5\%$ of its response in the non-rigid component and has a central-to-outer amplitude ratio of approximately $1.53$. The low-frequency Shankar features, on the other hand, are also predominantly non-rigid despite being spatially extended. This illustrates that non-rigidity and texture localization are separate properties: neither criterion alone is sufficient for assigning a texture-associated internal resonance.

The S1 and S2 amplitude maps occupy similar parts of the texture, but this does not mean that they describe the same oscillation. Because the harmonic analysis retains the phase and polarization of the response, the complete complex fields can be compared through
\begin{equation}
\mathcal{O}_{12}=\frac{\left| \sum_{\mathbf{r}} \delta \mathbf{n}_{\mathrm{int},1}^{*} \left( \mathbf{r} \right) \cdot \delta \mathbf{n}_{\mathrm{int},2} \left( \mathbf{r} \right) \right|}{\sqrt{\sum_{\mathbf{r}} \left| \delta \mathbf{n}_{\mathrm{int},1} \left( \mathbf{r} \right) \right|^2 \sum_{\mathbf{r}} \left| \delta \mathbf{n}_{\mathrm{int},2} \left( \mathbf{r} \right) \right|^2}}.
\label{eq:SM_overlap}
\end{equation}

For S1 and S2 we obtain $\mathcal{O}_{12} \simeq 0.114$. Thus, although their scalar amplitude profiles have substantial spatial overlap, their complex vectorial responses are nearly orthogonal. This provides additional evidence that the two spectral peaks correspond to distinct dynamical responses.

\subsection{Relation to the normal mode spectrum}\label{sec:sm-normal-modes}

The modes discussed in the main text are identified from the time-dependent response of the relaxed textures rather than from an explicit diagonalization of the atomistic dynamical matrix. The weak broadband pulse excites several frequencies simultaneously, which are then resolved from the subsequent free evolution together with their spatial profiles. Similar time-domain approaches have been used to study the excitation spectrum of magnetic hopfions, including breathing and rotational modes that can hybridize~\cite{raftrey2021field,bo2021spin}.

We identify H1, S1, and S2 as texture-associated resonances because their frequencies remain stable under the different spectral analyses described above, the corresponding spectral features are absent from the undriven trajectories, and the spatially resolved harmonic response extracted from the driven trajectories shows a non-rigid component spatially associated with the textures. We therefore interpret these features as texture-associated internal resonances. Strict identification with individual normal modes would require
linearizing the atomistic equations around the relaxed structure and solving the resulting eigenvalue problem, as in conventional internal mode calculations for magnetic skyrmions~\cite{lin2014internal}.

For the Shankar skyrmion, the dominant texture-associated resonance S1 exhibits a spatial response centered on the texture together with a coherent breathing-like deformation. This behavior naturally connects with the finite-frequency dilation coordinate $\lambda \left( t \right)$ obtained from the continuum theory, and we therefore associate S1 with the corresponding breathing sector. Since the atomistic response is not explicitly projected onto a dilation eigenvector, we use the term ``breathing-like'' rather than assigning S1 to a pure breathing eigenmode.

For the hopfion, H1 provides the clearest texture-associated internal resonance. Additional features are present in the spectrum, but their spatial response contains different mixtures of internal, extended, and global rotation dynamics, and they are therefore not included in the comparison of Fig.~\ref{FFT}. The coexistence and hybridization of different internal and rotational excitations is consistent with previous hopfion spectroscopy~\cite{bo2021spin}, while spatially localized breathing dynamics has also been reported for magnetic hopfions~\cite{tejo2026sub}.

The corresponding real-time post-pulse dynamics are provided as Supplementary Movies. Supplementary Movie S1 shows the evolution of the Shankar skyrmion following the weak anisotropy pulse, while Supplementary
Movie S2 shows the corresponding evolution of the hopfion under the same driving protocol. Because the deformation produced by the weak $K_z^0=10^{-5}$ pulse is difficult to discern directly on the scale of the
full texture, Supplementary Movie S3 provides an additional Shankar trajectory using a stronger pulse, $K_z^0=10^{-3}$ and $\tau_p=1$, evolved for $10^6$ integration steps. This stronger excitation is included for visualization and makes the breathing-like deformation of the Shankar texture clearly visible; it is not used in the spectral or spatial mode analysis presented above.

\newpage

\section{\NoCaseChange{Magnoom script for the frustrated antiferromagnet}}\label{magnoom_script}

\begin{lstlisting}
# Magnoom version employed in these simulations: https://github.com/rre93/magnoom
# Commit: 3486022aa6edab5778d969b64323bbc2a5756945
# Original repository: https://github.com/n-s-kiselev/magnoom

# begin magnoom config

# BatchMode: 1

# Lattice
# ax: 1
# ay: 0
# az: 0
# bx: 0
# by: 1
# bz: 0
# cx: 0
# cy: 0
# cz: 1

# System size
# Na: 64
# Nb: 64
# Nc: 64

# Boundary conditions and exchange interactions
# Shells: 4
# BCa: 1
# BCb: 1
# BCc: 1

# NOTE: Magnoom uses the opposite exchange-sign convention to that
# adopted in the manuscript
# J1: -0.205 // hopfion; use -0.2 for the Shankar skyrmion
# J2: 0.0
# J3: 0.0
# J4: -0.05

# Anisotropy
# VKu1: 0.0 0.0 1.0 // z-th axis
# Ku1: 0.0
# Ku2: 0.0
# Kc: 0.0

# Initial state
# InitialStateSource: ovf
# InputFile: AFM_hopf // hopfion; use AFM_shan for the Shankar skyrmion

# Anisotropy pulse
# AnisotropyStep: 1 // 1: pulse on, 0: pulse off
# Ku1Initial: 0.00001 // easy-axis pulse
# Ku1Final: 0.0
# AnisotropyStepIteration: 10 // pulse duration: t_step*AnisotropyStepIteration

# Dynamics
# damping: 0.0001
# t_step: 0.1
# IntegrationScheme: SIB

# Recording
# Recording: 1
# RecIteration: 1
# MaxIteration: 250000
# RecordedMoment: neel

# OVF output
# SaveOVF: yes
# OVFRecIteration: 1000
# OutputFile: X_easy_axis_Kz_0p00001_P10_alpha_0p0001 // X: hopfion, shankar 
# NOTE: OVF files are saved every 1000 iterations as <OutputFile>_iteration_XXXXXX.ovf

# TableFile: X_easy_axis_Kz_0p00001_P10_alpha_0p0001.csv // X: hopfion, shankar
# NOTE: CSV columns: iteration, time, Lx, Ly, Lz, total energy, and Ku1

# end magnoom config

\end{lstlisting}

\putbib[Bib-3D-Skyrmions]

\end{bibunit}

\end{document}